\documentclass[10pt,a4j]{article}
\usepackage[dvips]{graphicx}
\usepackage{mathrsfs}
\usepackage[dvips]{color}
\usepackage{cite}
\usepackage{framed}
\definecolor{shadecolor}{gray}{0.80}
\DeclareMathVersion{chem}
\usepackage{amsmath,amsthm,amssymb}
\usepackage{wrapfig}
\SetSymbolFont{letters}{chem}{OT1}{cmr}{m}{n}

\begin{document}

\renewcommand{\labelitemi}{}
\renewcommand{\thefootnote}{$\dagger$\arabic{footnote}}

\begin{flushright}
\textit{Concentration Dependence of Excluded Volume Effects}
\end{flushright}
\vspace{2mm}

\begin{center}
\setlength{\baselineskip}{25pt}{\LARGE\textbf{Concentration Dependence of Excluded Volume Effects}}
\end{center}
\vspace{-6mm}
\begin{center}
\setlength{\baselineskip}{25pt}{\large\textbf{Polymer Solution as Inhomogeneous System}}
\end{center}
\vspace*{4mm}
\begin{center}
\large{Kazumi Suematsu} \vspace*{2mm}\\
\normalsize{\setlength{\baselineskip}{12pt} 
Institute of Mathematical Science\\
Ohkadai 2-31-9, Yokkaichi, Mie 512-1216, JAPAN\\
E-Mail: suematsu@m3.cty-net.ne.jp,  Tel/Fax: +81 (0) 593 26 8052}\\[8mm]
\end{center}

%%%%%%%%%%%%%%%%%%
\hrule
\vspace{0mm}
\begin{flushleft}
\textbf{\large Abstract}
\end{flushleft}
The concentration dependence of the excluded volume effects in polymer solutions is investigated. Through thermodynamic arguments for the interpenetration of polymer segments and the free energy change, we show that the disappearance of the excluded volume effects should occur at medium concentration. The result is in accord with the recent experimental observations.
\begin{flushleft}
\end{flushleft}
\hrule
\vspace{3mm}
\setlength{\baselineskip}{13pt}
%%%%%%%%%%%%%%%%%% Introduction
\section{Introduction}
Concentration dependence of the excluded volume effects is an unsolved problem in polymer physics. To date, only a few theoretical works have been developed on this issue. Quite recently, very reliable experiments with the small angle neutron scattering (SANS) have been performed by two research groups\cite{Westermann, Graessley}; they showed that the unperturbed coil dimensions are realized at medium concentration far below the melt state; the unperturbed state being retained over wide concentration range.

In this paper we reexamine the concentration dependence of the excluded volume effects on the basis of the classic thermodynamic theory. The starting point of our investigation is motivated by the suggestion of the scaling formula by Issacson and Lubensky\cite{Issacson} that a critical concentration may occur for $d=3$ somewhere\cite{Kazumi} between the dilution limit and the non-solvent state above which the excluded volume effects vanish. 

A polymer solution is intrinsically an inhomogeneous system of segment concentration, which is due to the fact that monomers are joined by chemical bonds (see Fig. 1). A theory must therefore take this fact into consideration. Along this line, first we will make minor amendment of the classic theory of the local free energy in order to apply the theory in a more rigorous manner to the disappearance problem. Then, on this basis, we will show that the excluded volume effects should really vanish at medium concentration.

%%%%%%%%%%%%%%%%%% Theory
\section{Theoretical}
Excluded volume effects\cite{Flory, Fixman, Doi, Tanaka, Edwards, deGennes, Muthukumar, Cloizeaux, Schaefer, Plischke, Wittmer} of polymer molecules have two different facets: One is the expansion of the polymer dimensions and the other is the repulsion between two polymer molecules. These two phenomena, however, can be understood by a single thermodynamic property, the osmosis, namely the spontaneous flow of solvent from a more dilute region to a more concentrated region. The solvent flowing into from the dilute region will expand polymer coils. If all segments are joined by covalent bonds, the osmosis simply leads to the expansion of the polymer coil, while if the segments consist of some polymer molecules, the expansion will necessarily lead to the separation of those molecules, which will be phenomenologically observed like the core-core repulsion between hard spheres. Thus the coil expansion of one molecule and the repulsive interaction between two molecules are the different manifestations of the same phenomenon. 

In order for the osmosis to occur, there must exists concentration fluctuation in the system. Without the concentration gradient between the inside and the outside of the coil, no excluded volume effects can occur. Our task is then to estimate this gradient and calculate the magnitude of the exclude volume effects as a function of polymer concentration.

In order to investigate whether the transformation of the excluded volume coil to the ideal one occurs at a certain concentration, one must begin by evaluating the potential energy change $\Delta F$ of the system under the concentration fluctuation. More specifically, one must investigate whether the condition $\Delta F\ll kT$ can be fulfilled at certain finite concentration. If this inequality is satisfied, the excluded volume effect will be hidden behind the thermal fluctuation and never be detected experimentally.
Our question is then ``To what extent must the concentration fluctuation diminish in order for the excluded volume effect to be screened out?''\cite{Edwards, Muthukumar}.

\subsection{Local Free Energy: Minor Amendment}
In this section we make minor amendment of the classic theory of the local free energy\cite{Flory}; the amendment is necessary to apply the theory in a more rigorous manner to real polymer solutions. As is well known, the classic theory for the local free energy does not take into consideration correctly the interaction between segments on the same chain, so that the classic theory is constructed on the basis of the \textit{pseudo} self-avoiding chains whose behavior is virtually equivalent to that of the ideal chain\cite{Doi}. This feature has been frequently criticized so far as a fundamental deficiency of the Flory theory.

Let us consider the three dimensional lattice where a site may be occupied by a segment or a solvent molecule. Following the standard lattice representation, the random occupation of sites is assumed. Let a site be surrounded by $z$ neighboring sites, and let there be $\delta x$ segments from different molecules in the volume element $\delta V$. The number of arrangements of those segments is
%%%%%%%%%%%%%%%%%% eq-1
\begin{equation}
\Omega_{\delta V}=\prod_{i=0}^{\delta x-1}(z_{i}-1)(1-f_{i})
\end{equation}
where $f_{i}$ is the probability that a given cell is occupied by the polymer segments when $i$ segments are already put in $\delta V$, so that $f_{i}=i/\delta n_{0}$ where $\delta n_{0}=\delta x+\delta n_{1}$, $\delta n_{0}$ denotes the total number of sites and $\delta n_{1}$ the number of solvent molecules in $\delta V$. $z_{i}$ is a special number introduced in this amendment and is defined by $0\le z_{i}\le z$. The physical meaning of $z_{i}$ is as follows:

\vspace*{1mm}
\begin{minipage}[t]{0.95\linewidth}
\setlength{\baselineskip}{13pt}
There is finite probability that a given segment overlaps with the other segments on the same chain (the multioccupation problem). Such unphysical conformations must properly be removed by subtracting from the total number $\prod_{i}(z-1)$ of feasible conformations. This is possible, because the number of conformations is enumerable in principle. The subtraction can be achieved simply by reducing $z$ to $z_{i}$, so that $\prod_{i}(z_{i}-1)$ represents the total number of self-avoiding conformations. To date the numerical value of $z_{i}$ is unfortunately unknown\cite{Cloizeaux}, which however is not essential for the present purpose, as is verified below.
\end{minipage}

\vspace*{1.7mm}
\noindent By eq. (1) the local entropy becomes
%%%%%%%%%%%%%%%%%% eq-2
\begin{equation}
\delta S=\delta S_{mixing}+\delta S_{0}= k\, \log\, \Omega_{\delta V}=k \left\{\sum_{i=0}^{\delta x-1}\log\, (z_{i}-1)+\log\, \frac{\delta n_{0}!}{\delta n_{0}^{\delta x}(\delta n_{0}-\delta x)!}\right\}
\end{equation}
Applying the Stirling formula to the above equation, we have
%%%%%%%%%%%%%%%%%% eq-3
\begin{equation}
\delta S_{mixing}+\delta S_{0}\cong -k \left\{\delta n_{1} \log\,v_{1}+\delta x-\sum_{i=0}^{\delta x-1}\log\, (z_{i}-1)\right\}
\end{equation}
where $\delta S_{0}=\delta S_{01}(\delta x=0)+\delta S_{02}(\delta n_{1}=0)$ represents entropy for respective pure components and $v_{1}=(\delta n_{0}-\delta x)/\delta n_{0}$ is the volume fraction of solvents. It is clear by eq. (3) that $\delta S_{01}(\delta x=0)=0$, and $\delta S_{02}(\delta n_{1}=0)=-k\left\{\delta x-\sum_{i=0}^{\delta x-1}\log\, (z_{i}-1)\right\}$. Hence we have
%%%%%%%%%%%%%%%%%% eq-4
\begin{equation}
\delta S_{mixing}=-k\, \delta n_{1} \log\,v_{1}
\end{equation}
which is exactly the Flory result. It turns out that all the self-avoiding terms are absorbed into the melting entropy, $\delta S_{02}(\delta n_{1}=0)$. Noteworthy is the fact that the mixing entropy $\delta S_{mixing}$ of the \textit{pseudo} self-avoiding chains and solvent is exactly equal to that of the genuine self-avoiding chains and solvent; only the standard state must be altered as
\begin{equation}
\text{pure } pseudo \text{ self-avoiding chains}\hspace{5mm} \Rightarrow \hspace{5mm} \text{pure } genuine \text{ self-avoiding chains}\notag. 
\end{equation}
This unexpected result does not appear to have been fully recognized by theorists up to present. A close investigation however tells us that the result is by no means surprising, because by eq. (4) $\delta S_{mixing}$ is a function of the solvent fraction alone, but independent of the conformational properties of chains.

Adding the enthalpy term, $\delta H_{mixing}=kT\chi\delta n_{1}v_{2}$ to the above equation, we have the formula of the local free energy in the volume element $\delta V$
%%%%%%%%%%%%%%%%%% eq-5
\begin{equation}
\delta F_{mixing}=\delta H_{mixing}-T\delta S_{mixing}=kT\left\{\log\,\left(1-v_{2}\right)+\chi v_{2}\right\}\delta n_{1}
\end{equation}
where $v_{2}=1-v_{1}$ is the volume fraction of the segments and $\chi$ the enthalpy parameter. Eq. (5) represents the free energy difference between the mixture of the self-avoiding chains and solvent, and the respective pure components. Eq. (5) has already been derived by Flory\cite{Flory}.

We realize that eq. (5) has deeper generality along with sound physical basis, and hence is applicable equally to the excluded volume problem in concentrated systems.

%%%%%%%%%%%%%%%%%% Fig-1
\begin{wrapfigure}[15]{r}{7cm}
\vspace*{-6mm}
\begin{center}
\includegraphics[width=6.5cm]{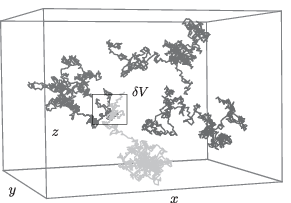}
\end{center}
\vspace*{2mm}
\setlength{\baselineskip}{10.5pt}{\small  Fig. 1: A snapshot of a polymer solution.}
\end{wrapfigure}

%%%%%%%%%%%%%%%%%% Subsection 2.1
\subsection{Expansion Factor $\alpha$ in Concentrated Solution}
Fundamental force of the coil expansion is the osmosis of solvent molecules from a more dilute region, $C_{valley}$, to a more concentrated region, $C_{hill}$ around the center of gravity of a coil, where $C$ denotes segment concentration. In order for the osmosis to occur for a given polymer coil, there must exists non-zero free energy difference between the inside of the coil ($C_{hill}$) and the outside ($C_{valley}$). If this circumstance is realized, positive molecular force from the outside region inevitably arises, creating the coil expansion.

The size of the expansion factor $\alpha$ is determined by the force balance between the osmosis and the retraction force due to rubber elasticity\cite{Flory} (see an alternative derivation in Appendix). The equilibrium condition is
%%%%%%%%%%%%%%%%%% eq-6
\begin{equation}
\partial F/\partial\alpha=\partial F_{osmotic}/\partial\alpha+\partial F_{elastic}/\partial\alpha=0
\end{equation}
By eq. (5) the free energy of mixing polymer segments and solvent molecules is written in the form:
%%%%%%%%%%%%%%%%%% eq-7
\begin{equation}
\Delta F_{M}=kT\int\left\{\log\,(1-v_{2})+\chi\hspace{0.3mm}v_{2}\right\}\delta n_{1}
\end{equation}
where the subscript 1 denotes the solvent, the subscript 2 the solute (polymer segment), and $\chi$ is the enthalpy parameter as mentioned above. Note that $\delta n_{1}=(1-v_{2})\hspace{0.3mm}\delta V/V_{1}$ with $V_{1}$ denoting the molecular volume of a solvent molecule. Substituting this into eq. (7), we have
%%%%%%%%%%%%%%%%%% eq-8
\begin{equation}
\Delta F_{M}=\frac{kT}{V_{1}}\int\left(1-v_{2}\right)\left\{\log\,(1-v_{2})+\chi\hspace{0.3mm}v_{2}\right\}\delta V
\end{equation}
Expanding the logarithmic term, we have
%%%%%%%%%%%%%%%%%% eq-9
\begin{equation}
\Delta F_{M}=\frac{kT}{V_{1}}\int\left\{-\left(1-\chi\right)v_{2}+\left(1/2-\chi\right)v_{2}^{2}+\frac{1}{6}v_{2}^{3}+\cdots\right\}\delta V
\end{equation}
Since we want to calculate the \textit{local} free energy difference, $\Delta F$, between the hill and the valley areas, let us write $\Delta F$ in the form:
%%%%%%%%%%%%%%%%%% eq-10
\begin{align}
\Delta F=&\Delta F_{M,hill}-\Delta F_{M,valley}\nonumber\\
=&\frac{kT}{V_{1}}\int\left\{-\left(1-\chi\right)(v_{hill}-v_{valley})+\left(1/2-\chi\right)(v_{hill}^{2}-v_{valley}^{2})+\frac{1}{6}\left(v_{hill}^{3}-v_{valley}^{3}\right)+\cdots\right\}\delta V
\end{align}
For $v_{2}\ll 1$ which is valid below the medium concentration under discussion, we may write
%%%%%%%%%%%%%%%%%% eq-11
\begin{equation}
\Delta F\cong \frac{kT}{V_{1}}\int\left\{-\left(1-\chi\right)(v_{hill}-v_{valley})+\left(1/2-\chi\right)(v_{hill}^{2}-v_{valley}^{2})\right\}\delta V
\end{equation}
Let $V_{2}$ denote the volume of a polymer segment and we have $v_{2}=V_{2}C$. For the Gaussian chain, with the equality $\delta V=\alpha^{3}d(x-a)d(y-b)d(z-c)=\alpha^{3}dxdydz$ in mind, we can write generally the segment concentration at a given point $(x,y,z)$ in the form
%%%%%%%%%%%%%%%%%% eq-12
\begin{equation}
C(x,y,z)=N\sum_{\{a,b,c\}}\left(\frac{\beta}{\alpha^{2}\pi}\right)^{3/2}e^{-\beta\{(x-a)^{2}+(y-b)^{2}+(z-c)^{2}\}}
\end{equation}
where $N$ is the number of the segments on a molecule, $\beta=3/2R_{g}^{2}$ with $R_{g}$ denoting the radius of gyration of an unperturbed chain and $(a,b,c)$ the coordinate of the center of gravity of a polymer molecule; so the summation represents the accumulation of segments emanating from a number of different polymers and therefore reflects the segment concentration at the point $(x,y,z)$. A graphical representation is given in Fig. 1 which shows that a polymer solution is the typical inhomogeneous system. In contrast to the homogeneous system of a monomer solution, there is wild fluctuation $\Delta C=C_{hill}-C_{valley}$ in the solution. $\Delta C$ is a function of $\alpha$; it rapidly decreases with increasing $\alpha$, because segments pervade more deeply the whole system as $\alpha$ increases. 

It is useful to extract the pre-factor from eq. (12) and recast it in the form
%%%%%%%%%%%%%%%%%% eq-12'
\begin{equation}
C(x,y,z)=N\left(\frac{\beta}{\alpha^{2}\pi}\right)^{3/2}G(x,y,z)\tag{12$'$}
\end{equation}
so that the quantity
%%%%%%%%%%%%%%%%%% eq-13
\begin{equation}
G(x,y,z)=\sum_{\{a,b,c\}}e^{-\beta\{(x-a)^{2}+(y-b)^{2}+(z-c)^{2}\}}
\end{equation}
is a function independent of $\alpha$. Now eq. (11) may be rewritten in the form
%%%%%%%%%%%%%%%%%% eq-14
\begin{equation}
\Delta F\cong NkT\frac{V_{2}}{V_{1}}\iiint\left\{-\left(1-\chi\right)\left(\frac{\beta}{\pi}\right)^{3/2}\left(G_{hill}-G_{valley}\right)+V_{2}N\left(1/2-\chi\right)\left(\frac{\beta}{\alpha\,\pi}\right)^{3}\left(G_{hill}^{\,2}-G_{valley}^{\,2}\right)\right\}dxdydz
\end{equation}
The first term of eq. (14) is, by eq. (13), independent of $\alpha$, so it vanishes by the differentiation with respect to $\alpha$. We have then
%%%%%%%%%%%%%%%%%% eq-15
\begin{equation}
\partial \Delta F_{osmotic}/\partial\alpha=-3N^{2}kT\frac{V_{2}^{\,2}}{\alpha^{4}V_{1}}\left(1/2-\chi\right)\left(\frac{\beta}{\pi}\right)^{3}\iiint\left(G_{hill}^{\,2}-G_{valley}^{\,2}\right)dxdydz
\end{equation}
Note that for any fluctuation model of polymer solutions, we can establish one-to-one correspondence between the $C_{hill}$ and the $C_{valley}$ by properly setting the defined spaces. Thus it is sufficient to take account of the elastic force per one molecule (this point will be discussed in Sec. 4). The classic work\cite{Treloar} showed that $\partial\Delta F_{elastic}/\partial\alpha=3kT\left(\alpha-1/\alpha\right)$. The force balance between the osmosis and the rubber elasticity is obtained by substituting these results in eq. (6), namely
%%%%%%%%%%%%%%%%%% eq-16
\begin{equation}
3N^{2}kT\frac{V_{2}^{\,2}}{\alpha^{4}V_{1}}\left(1/2-\chi\right)\left(\frac{\beta}{\pi}\right)^{3}\iiint\left(G_{hill}^{\,2}-G_{valley}^{\,2}\right)dxdydz=3kT(\alpha-1/\alpha)
\end{equation}
By rearrangement, we arrive at the expression (see an alternative derivation in Appendix):
%%%%%%%%%%%%%%%%%% eq-17
\begin{equation}
\alpha^{5}-\alpha^{3}=N^{2}\frac{V_{2}^{\,2}}{V_{1}}\left(1/2-\chi\right)\left(\frac{\beta}{\pi}\right)^{3}\iiint\left(G_{hill}^{\,2}-G_{valley}^{\,2}\right)dxdydz
\end{equation}
For the dilution limit, we have $G_{hill}\rightarrow e^{-\beta(x^{2}+y^{2}+z^{2})}$ and $G_{valley}\rightarrow 0$ and eq. (17) exactly reduces to the Flory result.

There are important implications in eq. (17): (i) A polymer coil is ideal ($\alpha=1$) at the theta point $(\chi=1/2)$ as is well established already, (ii) for a large $V_{1}$ limit, the coil is nearly ideal, and (iii) the fifth power rule of $\alpha$ is still valid, but (iv) the coil must also be ideal at the point of $G_{hill}=G_{valley}$ in which the concentration fluctuation disappears. 

\subsection{Excluded Volume \textit{u} in Concentrated Solution}
Prior to the derivation of the excluded volume $u$ in the concentrated solution, let us review briefly the classic work. Let a system contain only two polymer molecules. Consider the interpenetration of the two polymer molecules. Let a distance be $L$ between their centers of gravity. The chemical potential change $\left(\Delta F=\Delta H- T\Delta S\right)$ bringing these molecules close from $L=\infty$ to $L$ is
%%%%%%%%%%%%%%%%%% eq-18
\begin{equation}
\Delta F(L)=2kT\left(1/2-\chi\right)\left(V_{2}^{\,2}/V_{1}\right)\int \rho_{k}\rho_{\ell}\,dV
\end{equation}
where $\rho_{k}$ and $\rho_{\ell}$ are the respective segment density of the two polymer molecules $k$ and $\ell$ within the small volume element $\delta V$, which has the form:
%%%%%%%%%%%%%%%%%% eq-19
\begin{equation}
\rho_{k}=N\left(\frac{\beta}{\alpha^{2}\pi}\right)^{3/2}e^{-\beta\left(x_{k}^{2}+y_{k}^{2}+z_{k}^{2}\right)}
\end{equation}
Then the excluded volume $u$ can be calculated by the equation
%%%%%%%%%%%%%%%%%% eq-20
\begin{equation}
u=\int_{0}^{\infty}4\pi L^{2}\left(1-e^{-\Delta F(L)/kT}\right)dL
\end{equation}
The parameter $\chi$ is, according to the definition, positive for poor solvents but negative for good solvents. It follows that the interpenetration of the two polymer coils is strongly hindered in good solvents resulting in the molecular repulsion. We derive in the following the corresponding free energy change in the concentrated polymer solution.

%%%%%%%%%%%%%%%%%% Subsection 2.2.1
\subsubsection{Free Energy Change}
Suppose that a single polymer molecule $k$ is added to the concentrated polymer solution. There are two cases for the molecule to be located in the solution: one is the case where the molecule is put within the $C_{hill}$ region and the other is the case where it is put within the $C_{valley}$ region. The free energy difference between these two locations will correspond to the free energy difference between \textit{overlapping state} and \textit{non-overlapping state} in concentrated solutions. Now we can generalize the Flory excluded volume theory to include the concentrated system. The quantity in the small volume element $\delta V$ is
%%%%%%%%%%%%%%%%%% eq-21
\begin{multline}
\delta(\Delta F_{k,C})=\delta(\Delta F_{k,C_{hill}})-\delta(\Delta F_{k,C_{valley}})=kT\left(\delta V/V_{1}\right)\Big\{\left(1-\rho_{k}V_{2}-C_{hill}V_{2}\right)\log\left(1-\rho_{k}V_{2}-C_{hill}V_{2}\right)\\
-\left(1-C_{hill}V_{2}\right)\log\left(1-C_{hill}V_{2}\right)-\left(1-\rho_{k}V_{2}-C_{valley}V_{2}\right)\log\left(1-\rho_{k}V_{2}-C_{valley}V_{2}\right)\\+\left(1-C_{valley}V_{2}\right)\log\left(1-C_{valley}V_{2}\right)-2\chi\rho_{k}V_{2}^{2}\left(C_{hill}-C_{valley}\right)\Big\}
\end{multline}
Eq. (21) can be written in the series form:
%%%%%%%%%%%%%%%%%% eq-22
\begin{equation}
\delta(\Delta F_{k,C})=2kT\left\{\left(1/2-\chi\right)+\mathcal{O}\right\}\rho_{k}\left(C_{hill}-C_{valley}\right)\left(V_{2}^{2}/V_{1}\right)\delta V
\end{equation}
where the symbol $\mathcal{O}$ represents the higher terms of the series and a function of $C_{hill}$, $C_{valley}$ and $\rho_{k}$. Below the medium concentration, these terms are negligible and eq. (22) reduces to
%%%%%%%%%%%%%%%%%% eq-23
\begin{equation}
\delta(\Delta F_{k,C})\cong 2kT\left(1/2-\chi\right)\rho_{k}\left(C_{hill}-C_{valley}\right)\left(V_{2}^{2}/V_{1}\right)\delta V
\end{equation}
Let \textit{L} be a distance between the centers of $\rho_{k}$ and $C_{hill}$. Then $C_{valley}$ is also a function of \textit{L}, because the distance between $C_{hill}$ and $C_{valley}$ is fixed, on average, for the solution of a given concentration. Applying the equality $\delta V=\alpha^{3}dxdydz$ to eq. (23) and integrating over the defined spaces, we have the free energy difference as a function of \textit{L}
%%%%%%%%%%%%%%%%%% eq-24
\begin{equation}
\Delta F_{k,C}(L)= 2kT\left(1/2-\chi\right)\left(V_{2}^{2}/V_{1}\right)\alpha^{3}\iiint\rho_{k}\left(C_{hill}-C_{valley}\right)dxdydz
\end{equation}
In the limit of the infinite dilution, we have $C_{hill}\rightarrow\rho_{\ell}$ and $C_{valley}\rightarrow 0$, and we recover the classic equation (18). Using the Gaussian approximation, we may recast the above equation in the form:
%%%%%%%%%%%%%%%%%% eq-25
\begin{equation}
\Delta F_{k,C}(L)=2N^{2}kT\left(1/2-\chi\right)\left(V_{2}^{2}/V_{1}\right)\left(\frac{\beta}{\alpha\pi}\right)^{3}\iiint e^{-\beta\left(x_{k}^{2}+y_{k}^{2}+z_{k}^{2}\right)}\left(G_{hill}-G_{valley}\right)dxdydz
\end{equation}
\noindent where $G=\sum_{\{a,b,c\}}e^{-\beta\{(x_{\ell}-a)^{2}+(y_{\ell}-b)^{2}+(z_{\ell}-c)^{2}\}}$.
As the concentration fluctuation decreases, namely, $G_{hill}-G_{valley}\rightarrow 0$, then $\Delta F_{k,C}\rightarrow 0$. And it follows from eq. (20) that $u\rightarrow 0$. The excluded volume disappears in parallel with the disappearance of the fluctuation.

%%%%%%%%%%%%%%%%%% subsection-1.2
\section{Solution According to Lattice Model}
\subsection{Evaluation of Fluctuation}

By the equations (17) and (25), we have a general statement that the excluded volume effects are a strong function of the concentration fluctuation. This is a very natural conclusion, because the osmosis can occur only in the presence of the concentration gradient. Our task is then to evaluate the local concentration gradient around a given polymer coil.

Eqs. (17) and (25) are formal solutions not easy to solve, since the relative coordinates $(a,b,c)$ of individual molecules can not be specified in real solutions; moreover the boundary conditions of the integral terms in eqs. (17) and (25) are not clear. In this sense, no rigorous calculation seems possible. However it is possible to extract essential features of the equations by making use of the lattice model. In this paper we show the solution for eq. (17) only, since eq. (25) is more complicated and requires heavy calculation that seems to exceed the ability of computers available.

%%%%%%%%%%%%%%%%%% Fig-2
\begin{figure}[h]
\begin{center}
\begin{minipage}[t]{0.45\textwidth}
\begin{center}
\includegraphics[width=5.7cm]{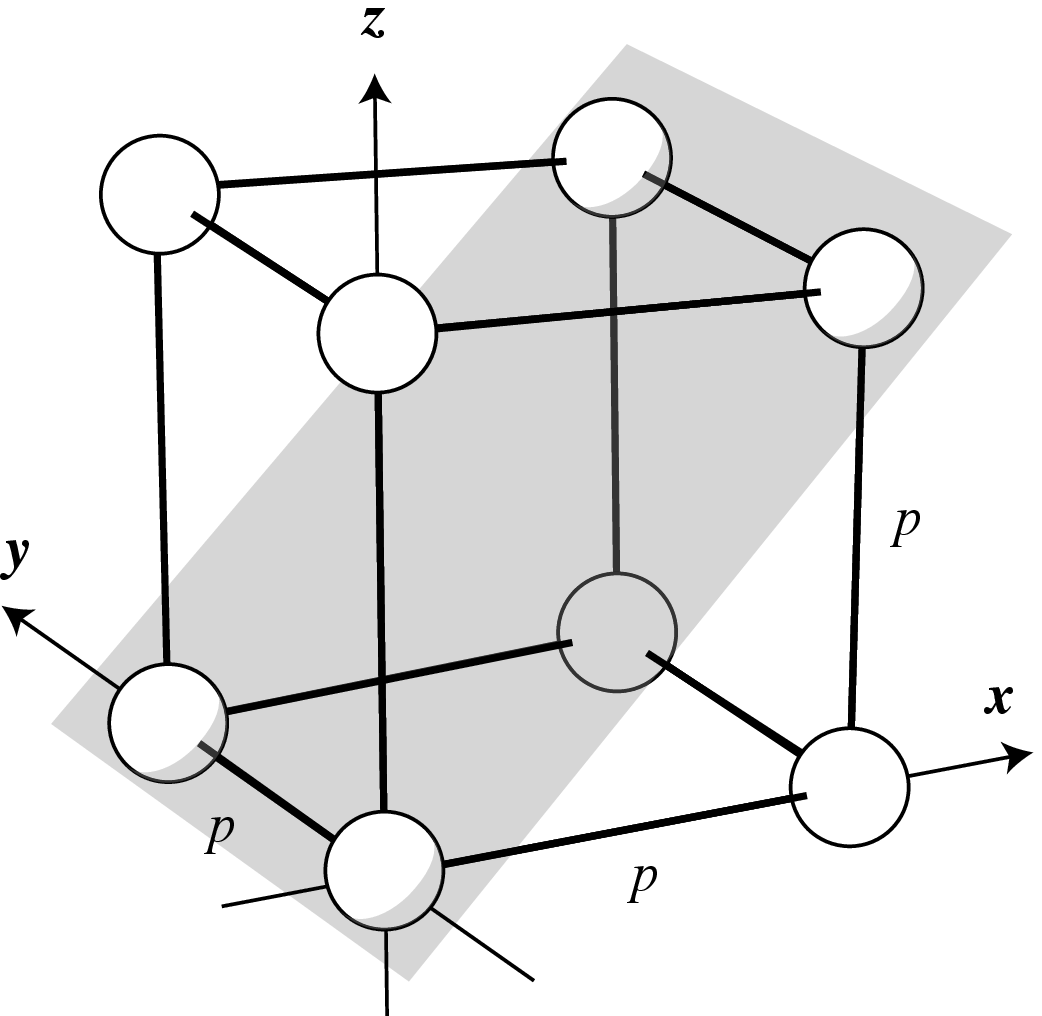}
\end{center}
\vspace*{0mm}
\setlength{\baselineskip}{10.5pt}{\small  Fig. 2: Representation of the simple cubic lattice: Edges are occupied by polymer molecules having the Gaussian distribution. The $C$ maximum and the minimum points lie on the $z=x$ plane.}
\end{minipage}
\hspace{10mm}
%%%%%%%%%%%%%%%%%% Fig-3
\begin{minipage}[t]{0.45\textwidth}
\begin{center}
\includegraphics[width=6.7cm]{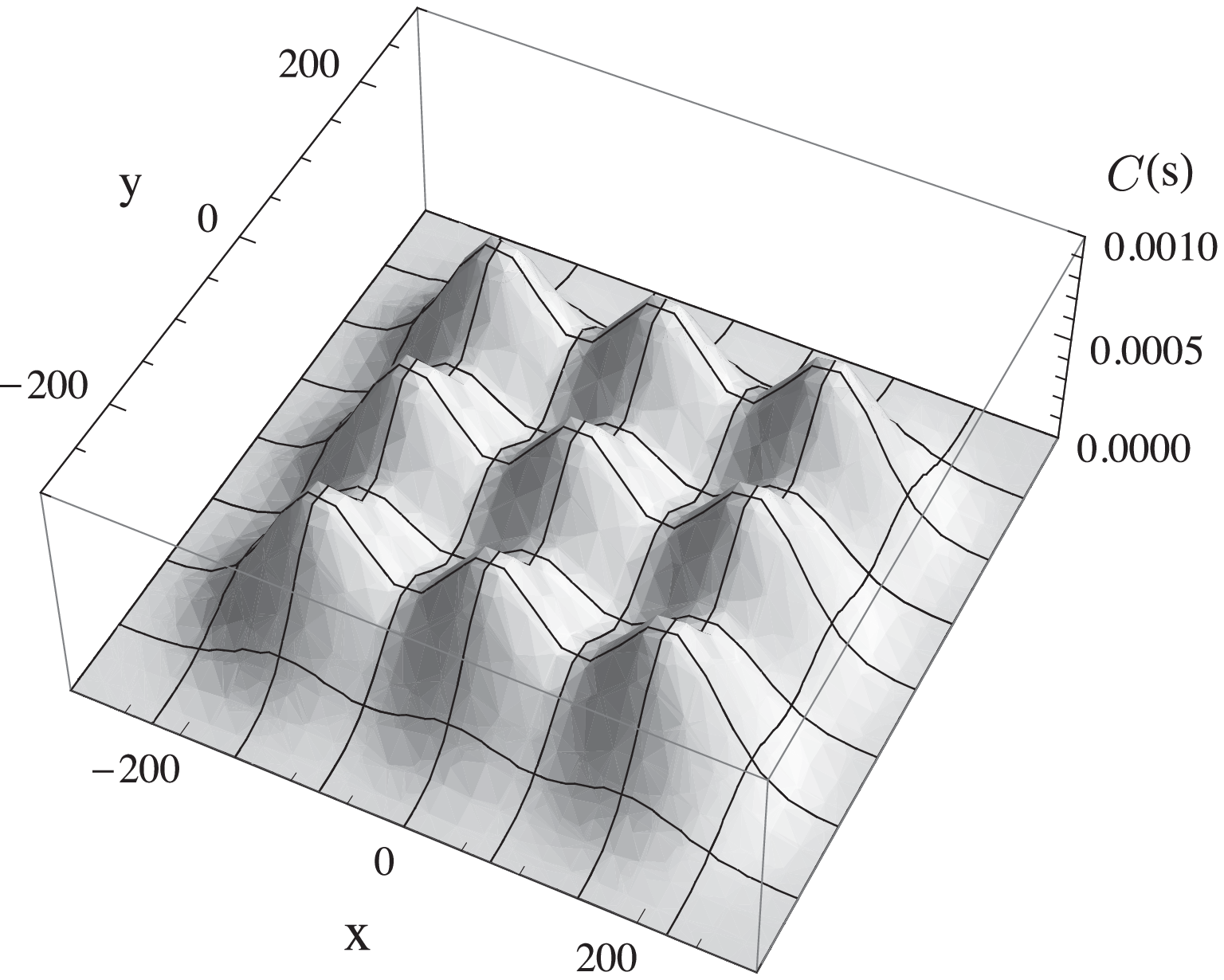}
\end{center}
\vspace*{0mm}
\setlength{\baselineskip}{10.5pt}{\small  Fig. 3: Representation of the concentration fluctuation in the solution of Gaussian polymers arranged on the simple cubic lattice. A peak represents the $C_{hill}$ and a bottom the $C_{valley}$: the center of a peak corresponds to the center of gravity of a chain.}
\end{minipage}
\end{center}
\end{figure}

Consider the simple cubic lattice with the unit lengths $(p\times p\times p)$. Polymer molecules are arranged on every sites with the Gaussian segment distribution. The relative coordinates of the molecules have then integer values of the form: $(a,b,c)=(ip,jp,kp)$ ($i,j,k=-\infty,\cdots, -1, 0, 1,\cdots,+\infty$). The segment number density $(/\text{\AA}^{3})$ at a given coordinate $(x,y,z)$ can be calculated by the simple summation of equal steps:
%%%%%%%%%%%%%%%%%% eq-26
\begin{align}
C(x,y,z)=&N\hspace{-2mm}\sum_{(i,j,k)=-\infty}^{+\infty}\hspace{-1mm}\left(\frac{\beta}{\alpha^{2}\pi}\right)^{3/2}e^{-\beta\left\{(x-ip)^{2}+(y-jp)^{2}+(z-kp)^{2}\right\}}\nonumber\\
=&N\left(\frac{\beta}{\alpha^{2}\pi}\right)^{3/2}{G}(x,y,z)
\end{align}
To find the maximum and the minimum points, differentiate eq. (26) to yield
%%%%%%%%%%%%%%%%%% eq-27
\begin{equation}
dC=\left(\frac{\partial C}{\partial x}\right)_{y,z}dx+\left(\frac{\partial C}{\partial y}\right)_{z,x}dy+\left(\frac{\partial C}{\partial z}\right)_{x,y}dz
\end{equation}
Since $x$, $y$ and $z$ are independent of each other, the solution for the equality $dC=0$ must satisfy
%%%%%%%%%%%%%%%%%% eq-28
\begin{equation}
\left(\frac{\partial C}{\partial x}\right)_{y,z}=\left(\frac{\partial C}{\partial y}\right)_{z,x}=\left(\frac{\partial C}{\partial z}\right)_{x,y}=0
\end{equation}
Here
%%%%%%%%%%%%%%%%%% eq-29
\begin{equation}
\left(\frac{\partial C}{\partial x}\right)_{y,z}=const\sum_{i=-\infty}^{+\infty}(x-i\,p)\,e^{-\beta\left\{(x-ip)^{2}+(y-jp)^{2}+(z-kp)^{2}\right\}}=0
\end{equation}
Obviously $x=i\,p$ satisfies eq. (29). The other solutions are $x=(1/2+\ell)\,p$ $(\ell=-\infty,\cdots, -1, 0, 1,\cdots,+\infty)$. From these, we have
%%%%%%%%%%%%%%%%%% eq-30
\begin{align}
C_{max}&=\{x=i\,p,\,y=j\,p,\,z=k\,p\}\nonumber\\
C_{min}&=\{x=(1/2+\ell)\,p,\,y=(1/2+m)\,p,\,z=(1/2+n)\,p\}
\end{align}
\noindent where $\ell, \, m, \,n=-\infty,\cdots, -1, 0, 1,\cdots,+\infty$. Thus $C_{max}$ and $C_{min}$ lie on $z=x+np$ planes. In Fig. 2, an example $(z=x$ plane) of those planes is illustrated. The mean number density of segments is directly calculated by the equation:
%%%%%%%%%%%%%%%%%% eq-31
\begin{equation}
\bar{C}=\frac{N}{p^{3}}\hspace{1cm} (/\text{\AA}^{3})
\end{equation}
so that the mean polymer volume fraction $\bar{\phi}$ is
%%%%%%%%%%%%%%%%%% eq-32
\begin{equation}
\bar{\phi}=\frac{N}{p^{3}}V_{2}
\end{equation}
An important quantity to be evaluated is the integral term appearing in the final expressions (17); this measures the 3-dimensional density fluctuation in the system. We put
%%%%%%%%%%%%%%%%%% eq-33
\begin{equation}
J_{\alpha}=\iiint\left(G_{hill}^{\,2}-G_{valley}^{\,2}\right)dxdydz
\end{equation}
As can be seen from Figs. 1-3, the $G_{hill}$ and the $G_{valley}$ areas are discontinuous; its domains can be partitioned into a number of intervals. By the lattice symmetry, we may define the intervals as $[-p/4,p/4]$ for $G_{hill}$ for each axis and $[p/4,3p/4]$ for $G_{valley}$. This choice is simply a matter of convenience for numerical calculation. Then the above integral may be specified as
%%%%%%%%%%%%%%%%%% eq-34
\begin{equation}
J_{\alpha}=\iiint_{-p/4}^{p/4}G^{\,2}\,dxdydz-\iiint_{p/4}^{3p/4}G^{\,2}\,dxdydz
\end{equation}

We show in Fig. 4-a the concentration fluctuation on the $x=y=z$ line that goes through the centers of the $C_{hill}$ and $C_{valley}$. The curves were calculated according to eq. (26) for $\alpha=1$ as a function of the mean polymer volume fraction $\bar{\phi}$, modeling polymethyl methacrylate (PMMA: $N=1000$) solutions. We have assumed the size of the segment to be equal to that of the repeating unit so that $V_{2}=140$ $\text{\AA}$. The light-brown peak with $\bar{\phi}=0$ represents the isolated polymer molecule in the dilution limit. As one can see, the concentration fluctuation decays rapidly with increasing $\bar{\phi}$ and vanishes for $\bar{\phi}\gtrsim 0.2$.

In Fig. 4-b, the quantity $J_{\alpha}/N$ is plotted as a function of $\bar{\phi}$ and  \textit{N}. The dotted line shows the result for $N=1000$ and the solid line the result for $N=5000$. The function $J_{\alpha}/N$ also decreases strongly with increasing concentration and vanishes at medium concentration.

%%%%%%%%%%%%%%%%%% Fig-4
\begin{center}
\hspace{-8mm}
\begin{minipage}[t]{0.95\textwidth}
\vspace*{0mm}
\begin{center}
\includegraphics[width=17cm]{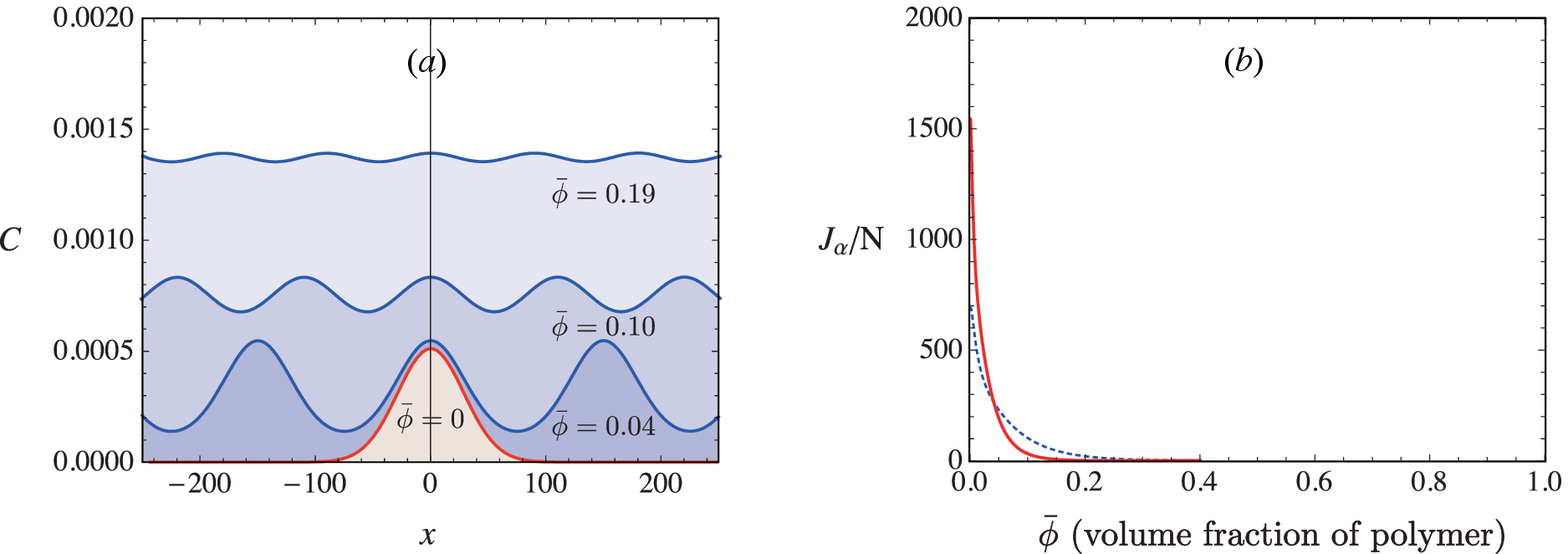}
\end{center}
\vspace*{-2.5mm}
{\setlength{\baselineskip}{10.5pt}{\small  Fig. 4: Concentration fluctuation in polymethyl methacrylate solutions ($V_{2}=140$). (a) Fluctuation of $C$ on the $x=y=z$ line (see also Fig. 2) as a function of the volume fraction $\bar{\phi}$ of the polymer ($N=1000$). (b) Numerical solution of $J_{\alpha}/N$ as a function of $\bar{\phi}$; $J_{\alpha}$ was calculated according to eq. (34) together with eq. (32); dotted line ($\textcolor{black}{\cdots}$): $N=1000$ and solid line ($\textcolor{black}{-}$): $N=5000$.}}
\end{minipage}%%%%%%%%%%%%%%%%%% Subsection 3.2
\subsection{Expansion Factor $\alpha$ as a Function of Concentration}
\end{center}
\vspace*{2mm}

With the help of the estimation of $J_{\alpha}$, we can now solve eq. (17) as a function of the segment concentration. Employed parameters are listed in Table 1. The enthalpy parameter $\chi$ is hard to estimate. Here we use the values of $\chi=0.3$ for PMMA$-$chloroform and $\chi=0.2$ for PE$-n$-nonadecane which is an assumed value at 150$^{\,\circ}$C by the extrapolation of the HandBook data\cite{Barton}.

The calculation results are illustrated in Figs. 5 and 6 to be compared with the observed points by Cheng, Graessley and Melnichenko\cite{Graessley} and those by Westermann, Willner, Richter and Fetters\cite{Westermann}. As one can see, the expansion factor $\alpha$ drops strongly with increasing polymer concentration and falls to the unperturbed value at the medium concentration in accordance with the experimental observations\cite{Westermann, Graessley}; the results are also consistent with the prediction of the Issacson-Lubensky scaling formula\cite{Issacson, Kazumi}.

According to the present calculation (Figs. 5 and 6), it is found that the essential features of the two systems of PMMA and PE are very alike: both of the systems show the swollen-to-unperturbed coil transition at the medium concentration. Comparing the two systems, it is found also that the coil expansion is less pronounced in the PE$-n$-nonadecane system than in the PMMA$-$chloroform system. The reason can be found by inspecting eq. (17); it simply comes from the special combination of the polymer and the solvent, namely $n$-nonadecane with the large molecular volume $V_{1}$ and polyethylene with the small segment volume $V_{2}$. According to eq. (17), such a combination necessarily depresses the expansion factor $\alpha$ to a lower level.

%%%%%%%%%%%%%%%%%% Table 1
\begin{table}[!htb]
\vspace{0mm}
\caption{Basic parameters of polymer solutions\cite{Graessley,Westermann}}
\begin{center}
\vspace*{-1mm}
\begin{tabular}{l l c r}\hline\\[-1.5mm]
& \hspace{13mm}parameters\footnotemark & notations & \hspace{4mm}values\,\, \\[2mm]
\hline\\[-1.5mm]
polymethyl methacrylate (PMMA) & volume of a solvent (CHCl$_{3}$) & $V_{1}$ & \hspace{5mm}$134\,\text{\AA}^{3}$\\[1.5mm]
& volume of a segment (C$_{5}$O$_{2}$H$_{8}$) & $V_{2}$ & \hspace{5mm}$140\,\text{\AA}^{3}$\\[1.5mm]
& degree of polymerization & $N$ & \hspace{5mm}5900\,\,\,\,\,\,\,\,\\[1.5mm]
& Flory characteristic ratio & C$_{F}$ & \hspace{5mm}9.2\,\,\,\,\,\,\,\,\\[1.5mm]
& mean bond length & $\bar{\ell}$ & $\hspace{5mm}1.56\text{\AA}$\,\,\,\\[1.5mm]
& enthalpy parameter (25$^{\,\circ}$C) & $\chi$ & \hspace{5mm}$0.3$\,\,\,\,\,\,\,\,\\[2mm]
\hline\\[-1.5mm]
polyethylene (PE) & volume of a solvent ($n$-C$_{19}$H$_{40}$) & $V_{1}$ & \hspace{5mm}$569\,\text{\AA}^{3}$\\[1.5mm]
& volume of a segment (C$_{2}$H$_{4}$) & $V_{2}$ & \hspace{5mm}$49\,\text{\AA}^{3}$\\[1.5mm]
& degree of polymerization & $N$ & \hspace{5mm}1000\,\,\,\,\,\,\,\,\\[1.5mm]
& Flory characteristic ratio & C$_{F}$ & \hspace{5mm}7.7\,\,\,\,\,\,\,\,\\[1.5mm]
& mean bond length & $\bar{\ell}$ & $\hspace{5mm}1.54 \,\text{\AA}$\,\,\\[1.5mm]
& enthalpy parameter (150$^{\,\circ}$C) & $\chi$ & \hspace{5mm}$0.2$\,\,\,\,\,\,\,\,\\[2mm]
\hline\\[-6mm]
\end{tabular}\\[6mm]
\end{center}
\end{table}

%%%%%%%%%%%%%%%%%% Fig-5
\begin{figure}[!htb]
\begin{center}
\begin{minipage}[t]{0.45\textwidth}
\begin{center}
\includegraphics[width=7.5cm]{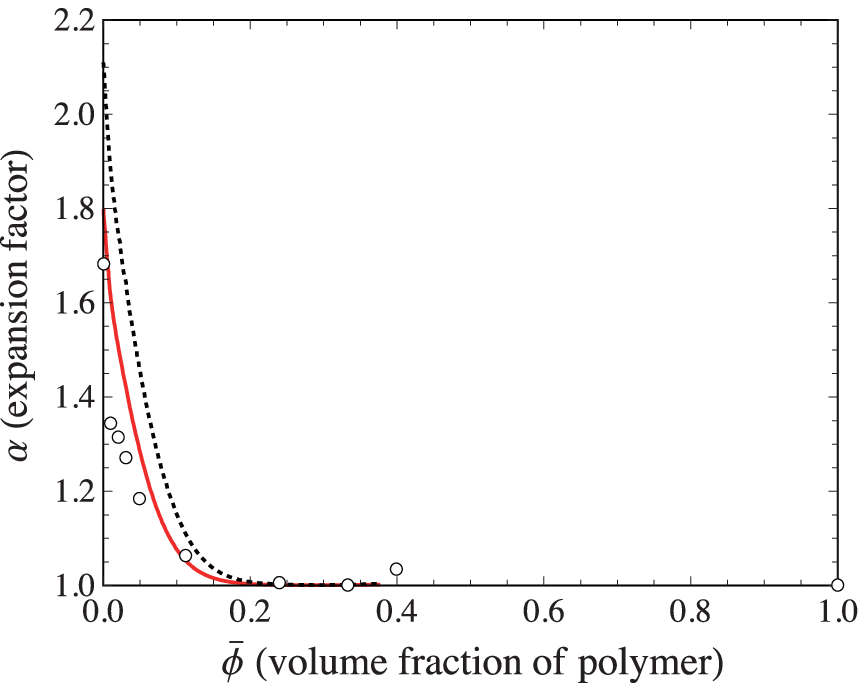}
\end{center}
\vspace*{0mm}
\setlength{\baselineskip}{10.5pt}{\small  Fig. 5: Expansion factor vs $\bar{\phi}$ plot for PMMA$-$CHCl$_{3}$. Solid line (\textcolor{black}{$-$}): theoretical line by eq. (17) for $\chi=0.3$. Dotted line ($\cdots$): theoretical line by eq. (17)  for $\chi=0$. Open circles ($\circ$): observed points by Cheng, Graessley and Melnichenko.}
\end{minipage}
\hspace{10mm}
%%%%%%%%%%%%%%%%%% Fig-6
\begin{minipage}[t]{0.45\textwidth}
\begin{center}
\includegraphics[width=7.5cm]{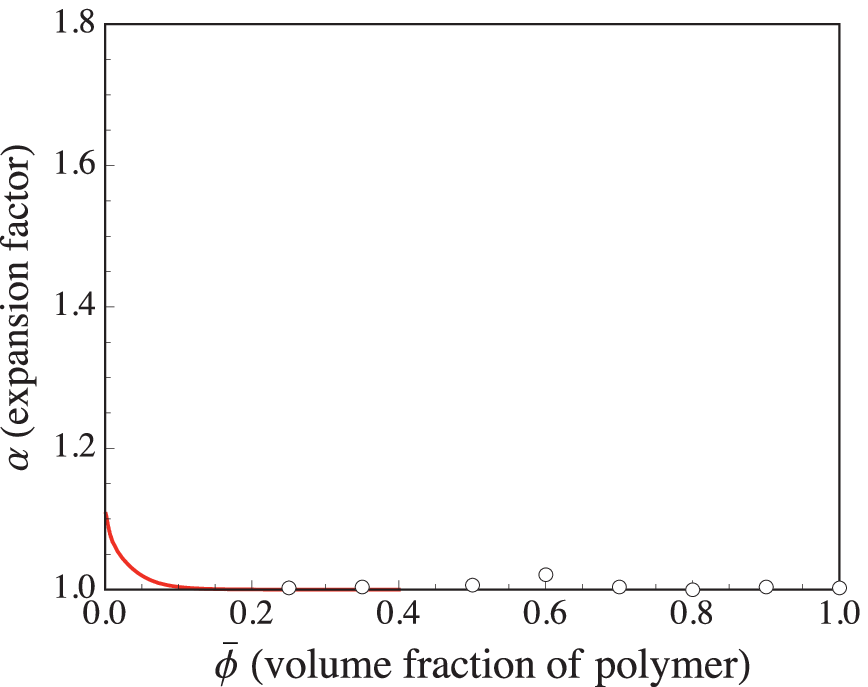}
\end{center}
\vspace*{0mm}
\setlength{\baselineskip}{10.5pt}{\small  Fig. 6: Expansion factor vs $\bar{\phi}$ plot for PE$-n$-C$_{19}$H$_{40}$. Solid line (\textcolor{black}{$-$}): theoretical line by eq. (17) for $\chi=0.2$. Open circles ($\circ$): observed points by Westermann, Willner, Richter, and Fetters.}
\end{minipage}
\end{center}
\end{figure}
\footnotetext{\ The unperturbed size is calculated by the equation: $R_{g}^{2}=\frac{1}{6}C_{F}N\xi\hspace{0.2mm}\bar{\ell}^{2}$, where $\xi$ is the bond number per one repeating unit. The numerical estimation of $R_{g}$ is in accord with the observed values\cite{O'Reilly}.}

%%%%%%%%%%%%%%%%%% Discussion
\section{Discussion}
The lattice model is a strong approximation where the dynamic aspect of real solutions is completely neglected. However, there are two cases in which this static model is expected to be a good representation of real solutions: (i) the lattice model will predict correctly the behavior in dilute solutions where polymer coils are, on average, far apart from each other so that the interaction among chains is weak; (ii) the model will predict correctly the location of the swollen-to-unperturbed coil transition at which the concentration fluctuation disappears. The agreement with the recent observations supports this reasoning.

One of the important questions throughout the present work was whether one can put the equality between the l.h.s and the r.h.s. terms in eq. (16); i.e., whether one can equate the osmotic pressure due to the density gradient exactly with the retractive force of one whole chain. To examine this problem, we must take the following fact into consideration: the osmotic phenomena within polymer coils are very different from the ordinary osmosis\cite{Glasstone}. The ordinary osmosis occurs between two systems separated by the semipermeable membrane, in which solvent molecules diffuse from a dilute system to a concentrated system, thus lowering the solute concentration in the concentrated system, whereas raising that in the dilute system. This process of the change of the solute concentrations on each side of the membrane takes place as a result of the volume change of the respective systems.

In contrast, there exists no semipermeable membrane in polymer solutions so that no volume change of the respective \textit{systems} can occur. The change of the concentrations in the two regions (dilute and concentrated) can occur only through the interchange of the space coordinates between polymer segments and solvent molecules, which leads to the coil expansion. The expansion of the polymer dimensions is therefore similar to the dissolution of a solid material into a solvent. The coil expansion resulting from the interchange of the coordinates tends to reduce the density fluctuation in the system, whereas the coil contraction will augment it. Only one way for the polymer solution to reduce its own fluctuation is to expand the polymer dimensions. Such adaptation of the coil dimensions should occur uniformly over the whole system because of the absence of the semipermeable membrane, thus validating the equality (16). Note that in the alternative derivation shown in Appendix, this problem on the equality (16) does not arise explicitly because of the delta function approximation for the segment density distribution.

There exists another question,``what do we mean by the unperturbed (or ideal) chain?'' To answer this question, let us return to the basic assumption of the present work. We have calculated the free energy difference $\Delta F$ by subtracting the free energy in the dilute region from that in the concentrated region. Thus the term ``unperturbed'' necessarily refers to the configuration under the condition that satisfies $C_{max}=C_{min}$. By the result of eq. (30) it corresponds to the configuration at $p=0$, namely the configuration at infinite concentration ($C\rightarrow\infty$) where all excluded volume effects should vanish rigorously as discussed earlier\cite{Kazumi}. Hence the term ``unperturbed'' defines a standard state for the configuration of a chain composed of mathematical dots and lines (with no volume and no thickness) and immersed in the true athermal \textit{solvent}. In that hypothetical limit, the segment should take the distribution of the form\cite{Ishihara}:
%%%%%%%%%%%%%%%%%% eq-35
\begin{equation}
W(s)=\frac{1}{N}\sum_{k=1}^{N}\left(\frac{9}{2\pi R^{2}\,\omega_{k}}\right)^{3/2}\exp\left(-\frac{9}{2R^{2}\,\omega_{k}}\,s^{2}\right)
\end{equation}
where $\omega_{k}=t_{k}^{3}+(1-t_{k})^{3}$, $t=k/N$ and $R$ is the end-to-end distance and $R^{2}=6Rg^{2}$. Usually this distribution is approximated by the corresponding Gaussian distribution as we have done in the text. Note that the above definition for the unperturbed chain is clearly different from that used in the classic theory\cite{Flory} where the standard state has been taken to be a pure polymer (melt state of self-avoiding chains). The two definitions are, however, virtually the same, since a chain obeys in effect the random flight statistics over all concentration range above the medium concentration, as we have seen in Figs. 5 and 6.

%%%%%%%%%%%%%%%%%% Conclusion
\section{Conclusion}
\begin{enumerate}
\item We have made a minor amendment for the local free energy, showing that the classic theory has deep generality and sound physical basis.
\item Making use of the thermodynamic arguments, we have developed the extended theory of the excluded volume effects, showing that the expansion factor $\alpha$ is a strong function of the concentration fluctuation $\rule[0.7mm]{6mm}{0.3pt}$ eq. (17). 
\item According to the solution of the lattice model, the fluctuation is maximum in the dilution limit, but decreases strongly as the polymer concentration increases, and vanishes at the medium concentration $\rule[0.7mm]{6mm}{0.3pt}$ Fig. 4. 
\item In parallel with the behavior of the fluctuation, the excluded volume effects manifest themselves most pronouncedly in the dilution limit, but decay rapidly with increasing concentration and vanish at the medium concentration.

The theoretical results are in good accord with the recent experimental observations\cite{Graessley,Westermann} $\rule[0.7mm]{6mm}{0.3pt}$ Figs. 5 and 6.
\end{enumerate}

%%%%%%%%%%%%%%%%%% Appendix
\section{Appendix}
\subsection*{An Alternative Derivation of Eq. (17)}
Let $W(s)$ be the unperturbed segment distribution around the center of gravity. We introduce the partition function for the excluded volume chain\cite{Fixman, Hermans}:
%%%%%%%%%%%%%%%%%% eq-A1
\begin{equation}
Z=\int_{s}W(s) \exp\left(-\frac{V(s)}{kT}\right)4\pi s^{2}ds\tag{A1}
\end{equation}
where $V(s)$ represents a potential function. Then the perturbed segment distribution can be formulated as
%%%%%%%%%%%%%%%%%% eq-A2
\begin{equation}
p(s)=\frac{1}{Z}W(s) \exp\left(-\frac{V(s)}{kT}\right)4\pi s^{2}\tag{A2}
\end{equation}
Making use of the Gaussian approximation for $W(s)$, we have
%%%%%%%%%%%%%%%%%% eq-A4
\begin{align}
\alpha^{2}=\frac{\langle s^{2}\rangle}{R_{g}^{2}}=\frac{\displaystyle\int_{s}s^{2}p(s)ds}{\displaystyle R_{g}^{2}\int_{s}p(s)ds}=\,\frac{\displaystyle\int_{t}\exp\left(-\frac{3}{2}t^{2}-\frac{V(t)}{kT}\right)t^{4}dt}{\displaystyle\int_{t}\exp\left(-\frac{3}{2}t^{2}-\frac{V(t)}{kT}\right)t^{2}dt}\tag{A3}
\end{align}
In eq. (A3), we have made the variable transformation: $t^{2}=s^{2}/R_{g}^{2}$. Independently we have another equality\cite{Hermans}:
%%%%%%%%%%%%%%%%%% eq-A4
\begin{equation}
\alpha^{2}=\frac{\displaystyle\int_{t}t\, \delta(t-\alpha)dt}{\displaystyle\int_{t}\frac{1}{t}\,\delta(t-\alpha)dt}\tag{A4}
\end{equation}
where $\delta(t-\alpha)$ signifies the delta function peaked at $t=\alpha$. If we identify eq. (A3) with eq. (A4), this amounts to making the approximation: $\delta(t-\alpha)\approx \exp\left(-\frac{3}{2}t^{2}-\frac{V(t)}{kT}\right)t^{3}$. Our remaining task is then only to evaluate the maximum point of the function, $\exp\left(-\frac{3}{2}t^{2}-\frac{V(t)}{kT}\right)t^{3}$, so that
%%%%%%%%%%%%%%%%%% eq-A5
\begin{equation}
\frac{d}{d t}\left.\log\left\{\exp\left(-\frac{3}{2}t^{2}-\frac{V(t)}{kT}\right)t^{3}\right\}\right|_{t=\alpha}=0\tag{A5}
\end{equation}
Here we use $\Delta F(\alpha\rightarrow t)$ in eq. (14) as the potential function $V(t)$. After some rearragement, we obtain
%%%%%%%%%%%%%%%%%% eq-A6
\begin{equation}
\alpha^{5}-\alpha^{3}=N^{2}\frac{V_{2}^{\,2}}{V_{1}}\left(1/2-\chi\right)\left(\frac{\beta}{\pi}\right)^{3}\iiint\left(G_{hill}^{\,2}-G_{valley}^{\,2}\right)dxdydz\tag{A6}
\end{equation}
which is just eq. (17) in the text.

%%%%%%%%%%%%%%%%%% References


\begin{thebibliography}{99}
\bibitem{Graessley} 
(a) W. W. Graessley. Polymer chain dimensions and the dependence of viscoelastic properties on concentration, molecular weight and solvent power. Polymer, \textbf{21}, 258 (1980).\\
(b) W. W. Graessley and R. C. Hayward. Excluded-Volume Effects in Polymer Solutions. 2. Comparison of Experimental Results with Numerical Simulation Data. Macromolecules, \textbf{32}, 3510 (1999).\\
(c) W. W. Graessley. Scattering by Modestly Concentrated Polymer Solutions. Macromolecules, \textbf{35}, 3184 (2002).\\
(d) G. Cheng, W. W. Graessley, and Y. B. Melnichenko. Polymer Dimensions in Good Solvents: Crossover from Semidilute to Concentrated Solutions. PRL, \textbf{102}, 157801 (2009).
\bibitem{Westermann} 
S. Westermann, L. Willner, D. Richter and L. J. Fetters. The evaluation of polyethylene chain dimensions as a function of concentration in nonadecane. Macromol. Chem. Phys., \textbf{201}, 500 (2000).
\bibitem{Issacson}  
(a) J. Issacson and T. C. Lubensky. Flory exponents for generalized polymer problems. J. Physique, Letters, \textbf{41}, L-469 (1980).\\
(b) T. C. Lubensky and J. Vannimenus. Flory approximation of directed branched polymers and directed percolation. J. Physique, Letters, \textbf{43}, L-377 (1982).
\bibitem{Kazumi} 
K. Suematsu. Recent Progress of Gel Theory: Ring, Excluded Volume, and Dimension.  Advances in Polymer Science, \textbf{156}, p-213 (Appendix) (2002).
\bibitem{Flory} 
P. J. Flory. Principles of Polymer Chemistry. Cornell University Press, Ithaca and London (1953).
\bibitem{Fixman} 
(a) M. Fixman. Excluded Volume in Polymer Chains. J. Chem. Phys., \textbf{23}, 1656 (1955).\\
(b) M. Fixman. Radius of Gyration of Polymer Chains. II. Segment Density and Excluded Volume Effects. J. Chem. Phys., \textbf{36}, 3123 (1962).
\bibitem{Edwards}
(a) S. F. Edwards. The theory of polymer solutions at intermediate concentration. Proc. Phys,  Soc., \textbf{88}, 265 (1966).\\
(b) S. F. Edwards. The size of a polymer molecule in a strong solution. J. Phys,  A: Math. Gen., \textbf{8}, 1670 (1975).
\bibitem{deGennes}  
P. G. de Gennes. Scaling Concepts in Polymer Physics, Cornell University Press, Ithaca and London (1979).
\bibitem{Muthukumar}
M. Muthukumar and S. F. Edwards. Extrapolation formulas for polymer solution properties. J. Chem. Phys., \textbf{76}, 2720 (1982).
\bibitem{Cloizeaux}
J. des Cloizeaux and G. Jannink. Polymers in Solution: Their modelling and structure. Clarendon Press, Oxford, Chapter 8 (1990).
\bibitem{Doi} 
M. Doi. Introduction to Polymer Physics. Clarendon Press, Oxford (1996).
\bibitem{Tanaka} 
F. Tanaka. Osmotic pressure of ring-polymer solutions. J. Chem. Phys., 87, 4201 (1987); Introduction to Physical Polymer Science, Shokabo Publishing Co., Ltd. (1994).
\bibitem{Schaefer}  
L. Sch\"afer. Excluded Volume Effects in Polymer Solutions: as Explained by the Renormalization Group. Springer-Verlag Berlin Heidelberg (1999).
\bibitem{Plischke}
M. Plischke and B. Bergersen. Equilibrium Statistical Physics. World Scientific, New Jersey, Chapter 8 (1999).
\bibitem{Wittmer}  
J. P. Wittmer, P. Beckrich, H. Meyer, A. Cavallo, A. Johner, and J. Baschnagel. Intramolecular long-range correlation in polymer melts: The segmental size distribution and its moments. Physical Review E, \textbf{76}, 011803 (2007).
\bibitem{Treloar} 
L. R. G. Treloar. The Physics of Rubber Elasticity. Oxford University Press (1958).
\bibitem{Barton}
A. F. M. Barton. CRC handbook of polymer-liquid interaction parameters and solubility parameters. CRC Press (1990).
\bibitem{O'Reilly}
J. M. O'Reilly, D. M. Teegarden, and G. D. Wignall. Small- and Intermediate-Angle Neutron Scattering from Stereoregular Poly(methyl methacrylate). Macromolecules, \textbf{18}, 2747 (1985).
\bibitem{Glasstone}
Samuel Glasstone. The Elements of Physical Chemistry. Macmillan \& Co Ltd, London (1956).
\bibitem{Ishihara}
(a) A. Ishihara. Probable Distribution of Segments of a Polymer Around the Center of Gravity. J. Phy. Soc. Japan, \textbf{5}, 201 (1950).\\
(b) P. Debye, and F. Bueche. Distribution of Segments in a Coiling Polymer Molecule. J. Chem. Phys., \textbf{20}, 1337 (1952).
\bibitem{Hermans}
(a) J.J. Hermans and J. Th. G. Overbeek. The dimensions of charged long chain molecules in solutions containing electrolytes. Rec. trav. chim., \textbf{67}, 761 (1948).\\
(b) T. B. Grimley. Equivalent Sphere Models of Real Chains. Trans. Faraday Soc., \textbf{55}, 681 (1959).
\end{thebibliography}
\end{document}